\documentclass[aps,prl,twocolumn,superscriptaddress]{revtex4}
\usepackage{amsfonts}
\usepackage{amsmath}
\usepackage{amssymb}
\usepackage{graphicx}
\usepackage{color}
\begin{document}
\title{Specific properties of supercooled water in light of  water anomalies}

%\date{\today }

\author{Victor Teboul}
\email{victor.teboul@univ-angers.fr}
\affiliation{Universit\' e d'Angers, Physics Department,  2 Bd Lavoisier, 49045 Angers, France}

\author{Ariadni P. Kerasidou}
\affiliation{Universit\' e d'Angers, Physics Department,  2 Bd Lavoisier, 49045 Angers, France}
\affiliation{ Institute of Nanoscience and Nanotechnology, NCSR "Demokritos", Athens, Greece}

\keywords{dynamic heterogeneity,glass-transition}
\pacs{64.70.pj, 61.20.Lc, 66.30.hh}

\begin{abstract}
We {\color{black} review} the effect of water anomalies on the properties of  low temperature water.
When supercooled, liquids dynamical properties change drastically. Supercooled liquids undergo an at least exponential  {\color{black} decrease of their diffusion coefficient} when temperature decreases while their structure merely doesn't change.   We {\color{black} discuss} how that still unexplained change of dynamical properties at low temperatures affect water differently from other liquids and what can be deduced from it.

%\vskip 0.5cm
%{\bf Main goals:}\\
%What are water specificities and how do these specificities act on the properties of supercooled water ?
%In this work we describe some of the water anomalies and their origins.

%\vskip 1cm
%Interest of that question:\\
%Water is one of the most important liquid on earth.  This is the liquid where biological processes take place.

\end{abstract}

\maketitle
\section{ Introduction}
%\subsection {a. Water at low temperatures}
%Water is ubiquitous in earth and living organisms ...
While of quite simple structure and composition, water is a strange liquid\cite{anomalies2}, displaying a number of anomalies\cite{anomalies,anomalies2}.
Water is also very important as the liquid where biological processes take place.
%Water while a simple three-atomic molecule leads to anomalous liquid and solid states.
The large number of anomalies relates to  the presence of hydrogen bonding that lead to the formation of a network organized liquid, 
and to various possible structures.
Merely a hundred anomalies have been reported for water, however the most striking anomalies are probably the existence of a maximum on the density and on the viscosity versus temperature. In other words, unlike in most other liquids, the density and viscosity of water do not evolve monotonously with temperature.
Three decades ago, to explain these behaviors, Poole et al. \cite{LDL2} postulated the existence of two different liquid states of water, a low density liquid (LDL)  and a high density liquid (HDL), with a critical point located at low temperature in the supercooled region \cite{LDL0,LDL1,LDL2,LDL3,LDL4,LDL5,LDL6,LDL7,LDL8,LDL9}. 

Below the homogeneous nucleation temperature $T_{H}=232 K$ water crystallizes rapidly \cite{valeria1,valeria2,valeria3,nucleus1,nucleus2}. Thus it is very difficult to access experimentally with bulk liquid water the region of temperature below $T_{H}$ and above the glass-transition $T_{g}=136 K$. Due to that difficulty, that region of temperature ($T_{H}>T>T_{g}$) has been called the 'no man's land'.
As the putative liquid-liquid transition between HDL and LDL is located inside the 'no man's land' it is difficult to observe experimentally and  the metastability of these two liquid states is still the subject of controversy \cite{David1,David2}.
We will nonetheless in this paper use that picture of coexisting different structures (polyamorphism) to understand how it fits or not the properties that we observe at low temperatures.

%\vskip0.5cm
%\subsection {b. The glass transition problem}

When cooled rapidly enough, liquids can remain in the liquid state below their melting temperature\cite{book0}. 
Usually, the more complex the molecule, the easier the crystallization can be avoided leading to that supercooled 'state'.
As discussed above, for water the crystallization always occurs when the temperature decreases below a limit temperature $T_{H}=232 K$, leading to the so called 'no-man's land' region of temperature that thus cannot be accessed experimentally for bulk supercooled water.
If the temperature can be further decreased, the viscosity of the liquid $\nu$ increases exponentially for Arrhenius (or strong) liquids like silica or even more rapidly for super-Arrhenius (or fragile) liquids like most molecular liquids and water. The viscosity $\nu$ and diffusion coefficient $D$ follow  laws of the form:
 \begin{equation}
 \nu= \nu_{0} exp(E_{a}/kT)
 \end{equation}
 and
 \begin{equation}
D= D_{0} exp(-E'_{a}/kT)
\end{equation}
With  activation energies $E_{a}(T)$ and $E'_{a}(T)$  that are approximately constant for strong liquids and increase for fragile liquids when temperature drops. 
Note that around the melting temperature $E_{a}(T) \approx E'_{a}(T)$ then they differ at lower temperature due to the breaking of the Stokes-Einstein law.
{\color{black} However as the trend of the variations of both quantities is relatively similar (if one is inverted) as shown in equations 1 and 2, we will use in this work sometimes the term viscosity and sometimes the diffusion coefficient, but only scarcely both, to avoid repetitions.}
Eventually, at low enough temperature the medium becomes so viscous that it behaves as a solid and  is called a glass.
The reason for that increase in viscosity that occurs without structural change is however still unknown\cite{anderson} and the object of active researches.
In the supercooled state, spontaneous transient cooperative motions, called dynamic heterogeneities, appear in the liquid and increase when the temperature drops.
The appearance of cooperative motions\cite{DH} that is also still not explained, is however expected as a signature of the approach of a phase transition\cite{book1}.

\vskip0.5cm
%{\color{black} \bf Connections of the anomalies with low temperature behavior}\\

\section{Models}

A variety of intermolecular potentials exist to model water\cite{SPC,SPCE,TIP5Pa,TIP5Pb,TIP5PE,POT1,POT2}. 
{\color{black}The comparison of the properties of water obtained using different potentials has been the subject of various works, see for example\cite{mm,TIP5Pb,SPC}.}
We list here the most important potentials for supercooled water simulations.
One of the simplest and most often used water potentials\cite{SPC,SPCE} is the SPC(E) potential. 
Due to its simplicity it is very efficient for simulations as it takes only three beads into account to model the three atoms molecule.
Unfortunately the diffusion is too large with that potential in comparison with experiments. 
The TIP5P(E) potential\cite{TIP5Pa,TIP5Pb,TIP5PE} is a little more complex and less efficient as it models water with $5$ beads instead of $3$,
but it leads to a better dynamics{\color{black} \cite{mm}}, a better structure and {\color{black}can induce} crystallization.
Coarse graining\cite{book2,CG1} is often used in simulations to increase drastically the efficiency of the calculations.
In biological systems a very large number of water molecules has to be used due to the large size of the biopolymers they surround.
Thus coarse graining of water is of large interest, but rather difficult to model due to the hydrogen bondings, polyamorphism and electrostatic interaction. Several coarse grained potential have however been proposed for water.
The Molinero's coarse grained potential\cite{POT2} that model water as an intermediate element between Carbon and Silicon, is very efficient and has been used extensively for low temperature simulations.

%\subsection{Treatment of long range interactions}
Long range electrostatic interactions, while screened\cite{structure,screening} by the presence of the surrounding molecules, have to be handled with some care.
While a few models use a simple cutoff for interactions at distances larger than $R_{cutoff}$, the most common ways of dealing with electrostatic interactions are the Ewald method and the Reaction field method.
{\color{black}The Ewald method uses the infinite number of replica of the simulation box (Born- Von Karman periodic conditions) to calculate the global interactions on each atom. To solve the infinite range calculation,  the long range part of the global potential is transformed into Fourier space where the calculation is short ranged. 
The Reaction field method approximates the long range interactions (defined as the interactions for distances larger than a chosen cutoff $R_{c}$) between each atoms, by the interaction with a continuous medium with  dielectric constant $\epsilon$.}

The two methods are included in most simulation codes. However for scientists using their own simulations programs the reaction field is much easier to implement and leads to faster simulations.
For the physics of the system, we expect the Ewald method to somehow increase the tendency for crystallization, while the reaction field will not affect that tendency. 
Notice that it is important in the reaction field method to apply the reaction field cutoff on the whole molecule (i.e. on the center of masses) to avoid charge fluctuations.
Note also that at low temperatures, finite size effects\cite{finite1,finite2} appear in supercooled liquids and modify their dynamics. Thus the simulation box has to be large enough to prevent from these effects to happen, and the lower the temperature, the larger the box must be chosen to avoid finite size artifacts.

In this paper we model the water molecular interactions with the TIP5PE potential\cite{TIP5PE} and the long range electrostatic interaction with the Reaction field method using a cutoff radius $R_{c}=9$\AA\ and an infinite dielectric constant for distances $r>R_{c}$.
The water molecule is modeled as a rigid body and we will focus our attention on  the center of masses behavior (that is also approximately the oxygen atoms behavior as the differences in the position of the center of masses and of the oxygen atom is quite small in water).
Our simulation box contains $2000$ water molecules in a cubic box with usual periodic conditions and is aged at the temperature of study during $10ns$ (for $T \geq 250 K$) or $20 ns$ (for $T \leq 240K$) before any recorded run.  

\section{Results and discussion}

We will show a few results from simulations of supercooled water before discussing them, together with previous works with the perspective of  a connection with the water anomalies.

%\subsection{ Anomalous properties above $T_{m}$}

It is possible to rationalize most of the anomalous properties of water from its two most important particularities:  The presence of hydrogen bounding leading to a network structure and the polyamorphism or equivalently the existence of several metastable structures.
We will now discuss the properties of supercooled water in relation to these two important particularities of water.

\vskip 0.5cm
\subsection{ Large super-Arrhenius behavior}

Water is a fragile liquid in Angell's classification, as its viscosity increases {\color{black}(diffusion decreases)} more than exponentially when the temperature drops below its melting temperature. 
%We display that behavior in Figure 3.
Most molecular liquids are fragile, however water is a very fragile liquid and the super-Arrhenius evolution of the viscosity with temperature is larger in water than in most liquids.
Fragility is usually related to the extent of cooperative motions as one explains the increase of the activation energy $E_{a}(T)$  by the need for an increasing number of molecules to cross energy barriers cooperatively for diffusion to take place.  
Consequently, the large super Arrhenius behavior of water suggests that cooperative motions are particularly large for water.

\begin{figure}
\centering
\includegraphics[height=6.5cm]{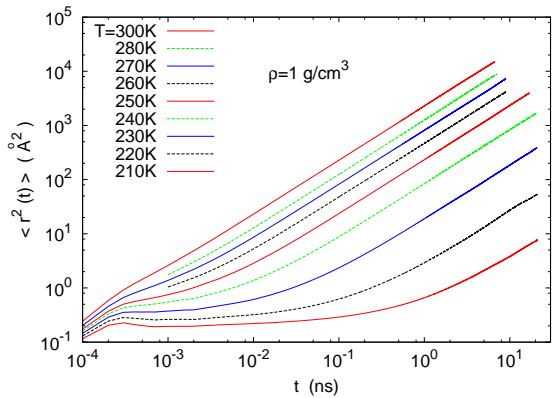}
\caption{(color online) Mean square displacement of the center of masses versus time for various temperatures at the water density $\rho=1g/cm^{3}$.
Note the three time regimes characteristic of supercooled liquids appearing below $T_{m}$.
For very short time scales we observe the ballistic regime with molecules still moving almost freely like in a gas, then for intermediate time scales the plateau regime arising due to the caging and reminiscent of the solid state, and eventually the diffusive regime when molecules escape the cage of their neighbor displaying a liquid behavior.
} 
\label{f1}
\end{figure}

\begin{figure}
\centering
\includegraphics[height=6.5cm]{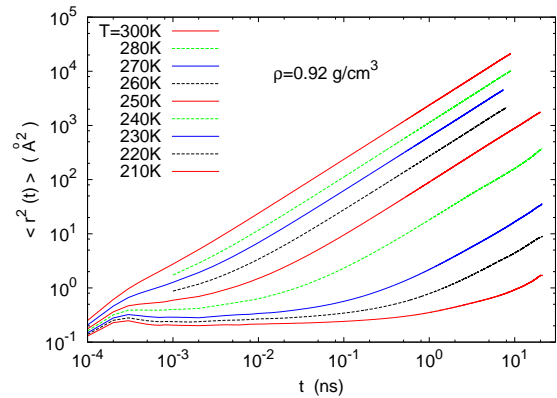}
\caption{(color online) As in Figure 1 but for a smaller water density $\rho=0.92 g/cm^{3}$.
Note that the dynamics is slower here than in Figure 1 for  $\rho=1 g/cm^{3}$. 
That is one of the strange particularities of supercooled water. %That behavior can be related to the non-monotonous evolution of the viscosity and density with the temperature above $T_{m}$, and to the existence of different possible structures (polyamorphism).
} 
\label{f2}
\end{figure}

We will now illustrate the evolution of the diffusion with temperature in supercooled water from the calculation of the mean square displacements and diffusion coefficients of oxygen atoms.
Figures \ref{f1} and \ref{f2} show the mean square displacement $<r^{2}(t)>$ evolution with temperature of supercooled water, respectively at the density of room temperature liquid water and slightly above the density of ice (at atmospheric pressure).
Below the melting temperature $T_{m}=273 K$ that corresponds approximately to the third curve from the top, the mean square displacements display the three characteristic time regimes of supercooled liquids, namely the ballistic time regime at short time scales (below $0.3 ps$), the plateau regime for intermediate time scales and the diffusive time regime for large time scales.
While the temperature decreases, the plateau time regime increases showing that the molecules need more time to escape the cages of their neighbors, leading to a slowing down of the dynamics and a decrease of the diffusion coefficient.
If we compare Figures  \ref{f1} and \ref{f2} that correspond to the densities $\rho=1 g/cm^{3}$ and $\rho=0.92 g/cm^{3}$ respectively, we see that the plateaus are larger for the same temperatures in Figure \ref{f2}. The diffusion is smaller (or viscosity larger)  for the smallest density.
That counterintuitive property is one of the water anomalies that the presence of several possible structures explains. 

\begin{figure}
\centering
\includegraphics[height=6.5cm]{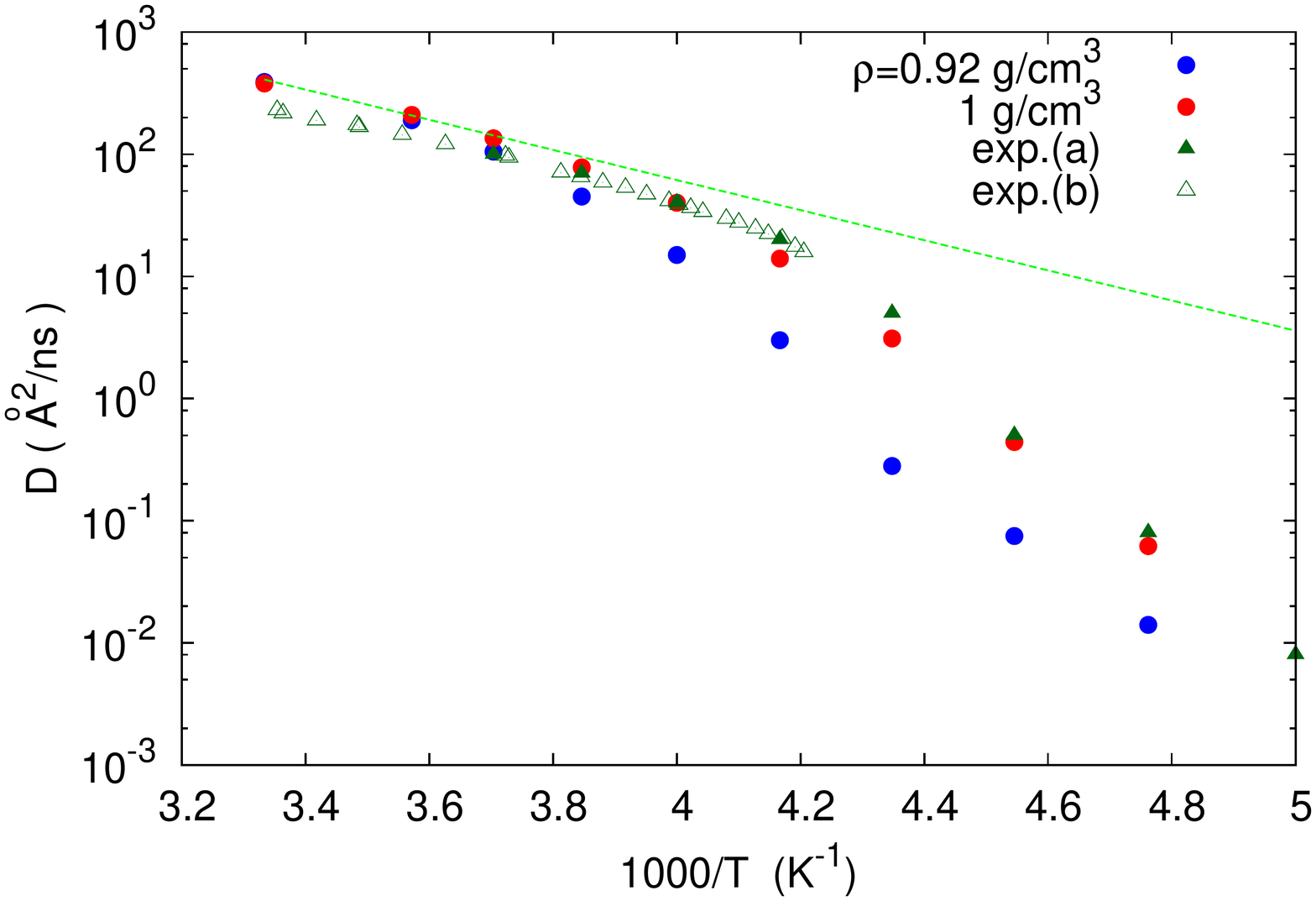}
\caption{(color online) Diffusion coefficient of oxygen atoms (or centers of masses of water molecules) versus the inverse of the temperature, at constant density.
The dashed line is the curve expected from an Arrhenius pure exponential evolution of the diffusion coefficient.
{\color{black} The triangles are experimental data  obtained (using the Wilson-Frenkel model) from ice growth rate experiments at low pressure (full triangles (a) \cite{exp}) or directly measured at ambient pressure (empty triangles (b) \cite{exp2}). Note that some of these data are inside the no man's land.
We expect the triangles to be located in between the blue and red circles as the density of water at ambient pressure decreases at low temperature. The model is thus slightly less diffusive than the experimental data at low temperatures.}
Note that the smallest density (blue circles, $\rho=0.92 g/cm^{3}$) induces the largest departure from the Arrhenius evolution, suggesting larger cooperative motions at low density, a particularity of water.
} 
\label{f3}
\end{figure}

Figure \ref{f3} shows the diffusion coefficient as a function of temperature for the two densities considered in our simulations.
The triangles correspond to recent experimental data\cite{exp} at ambient pressure.
The diffusion coefficient $D$ is here calculated from the relation:
 \begin{equation}
\lim_{t \to \infty}<r^{2}(t)>=6 D t
\end{equation}
We see on the Figure that the diffusion coefficient departs rapidly from the Arrhenius pure exponential law materialized by the green dashed  line.
This evolution shows that water is a very fragile liquid in that temperature range.
The activation energy $E_{a}(T)$ increases significantly when the temperature drops, suggesting the presence of large cooperative motions.
Understanding the activation energy as the energy that is necessary to overcome, to be able to move inside the environment, an increase of the activation energy means that several molecules have to overcome the one-molecule activation energy for the motion to be possible. 
As a result the increase of the activation energy  when the temperature drops reflects the average number of molecules involved in cooperative clusters at different temperatures.

\vskip 0.5cm
\subsection{ Effect of {\color{black} density or pressure}}

Increasing the pressure or the density leads to a decrease of the super-Arrhenius behavior\cite{pressure1,pressure2,pressure} in water.
This effect has been observed experimentally and with molecular dynamics simulations using various potentials.
We observe that effect in the diffusion coefficients of Figure \ref{f3}, and by comparison of the mean square displacements of Figures \ref{f1} and \ref{f2} {\color{black}for the two different densities studied}.
As the density is made smaller in Figure \ref{f3}, the diffusion departs more from the Arrhenius law displayed with a green line and consequently water becomes more fragile.

As the fragility is connected with the cooperative motions, we deduce that increasing the pressure (or the density) decreases the cooperative motions and accelerates the dynamics (i.e. decreases the viscosity) in supercooled water.
We interpret this effect from the pressure induced structural modifications in the liquid.
Increasing the pressure promotes the high density structure and as a result decreases the structural fluctuations that are responsible for the large cooperative motions in water.

\vskip 0.5cm
\subsection{ Large cooperative motions}

We will now use a more direct measure of the clusters of cooperative motions inside the liquid.
For that purpose we define the Non Gaussian parameter (NGP) $\alpha_{2}(t)$ as:

 \begin{equation}
 \alpha_{2}(t) = {{3 <r^{4}(t)>} \over{ 5<r^{2}(t)>^{2}}} -1
 \end{equation}

\begin{figure}
\centering
\includegraphics[height=6.5cm]{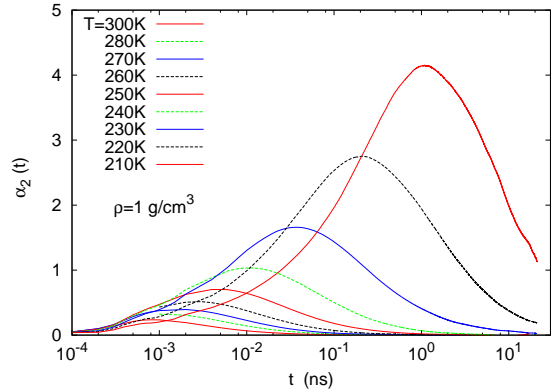}
\caption{(color online) Non Gaussian parameter $\alpha_{2}(t)$ versus time for various temperatures at constant density $\rho=1 g/cm^{3}$.
$\alpha_{2}(t)$ quantifies the deviation of the self Van Hove distribution function from a Gaussian. 
Consequently $\alpha_{2}(t)$  is associated to the appearance of cooperative motions that are at the origin of the Van Hove shape modification, and it is used as a measure of the intensity of these motions. 
} 
\label{f4}
\end{figure}

\begin{figure}
\centering
\includegraphics[height=6.5cm]{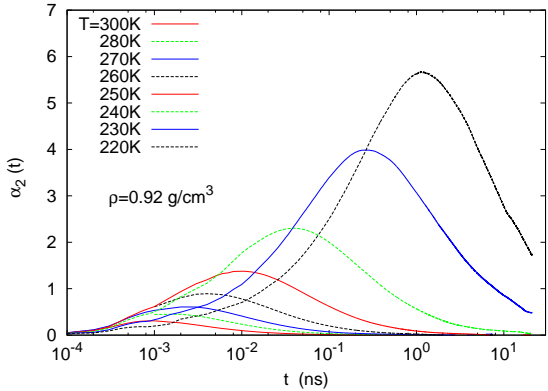}
\caption{(color online) As in Figure 4 but at a smaller density $\rho=0.92 g/cm^{3}$.
} 
\label{f5}
\end{figure}

When the temperature drops, the self part of the Van Hove correlation function\cite{bookmd} that represents the probability at time $t$ to find a molecule a distance $r$ apart its previous position at time zero, changes.
A tail develops for large $r$ values, that is the signature of cooperative motions that are larger than the average motions.
As a result the Van Hove is no longer a Gaussian.
The Non-Gaussian parameter is a measure of that departure from a Gaussian shape and consequently quantifies the cooperative motions.
Figures \ref{f4} and \ref{f5} show the Non Gaussian parameter evolution when the temperature drops for the two densities of our study.
The Non-Gaussian parameter is small at high temperature and progressively increases when the temperature drops.
Simultaneously the curves are shifted to larger time scales due to the decrease of the probability for a molecule to escape the cage of its neighbors.
The maximum of the curves corresponds to the plateau ending time regime for the mean square displacement.
For that time, some molecules are escaping the cages while over are still trapped.
In what follows we will call $t^{*}$  that characteristic time for which the NGP is maximum. This time has been shown in numerous works{\color{black} \cite{DH,DH0,DH1,DH2,DH3,finite1,bulk,silica}} to be a universal characteristic of  cooperative motions in supercooled liquids.
{\color{black} It was found to correspond to the characteristic time of string-like motions, that is the typical time for a molecule to replace another  in a string of mobile molecules\cite{DH,DH1,DH2,bulk,silica}.}

\begin{figure}
\centering
\includegraphics[height=6.5cm]{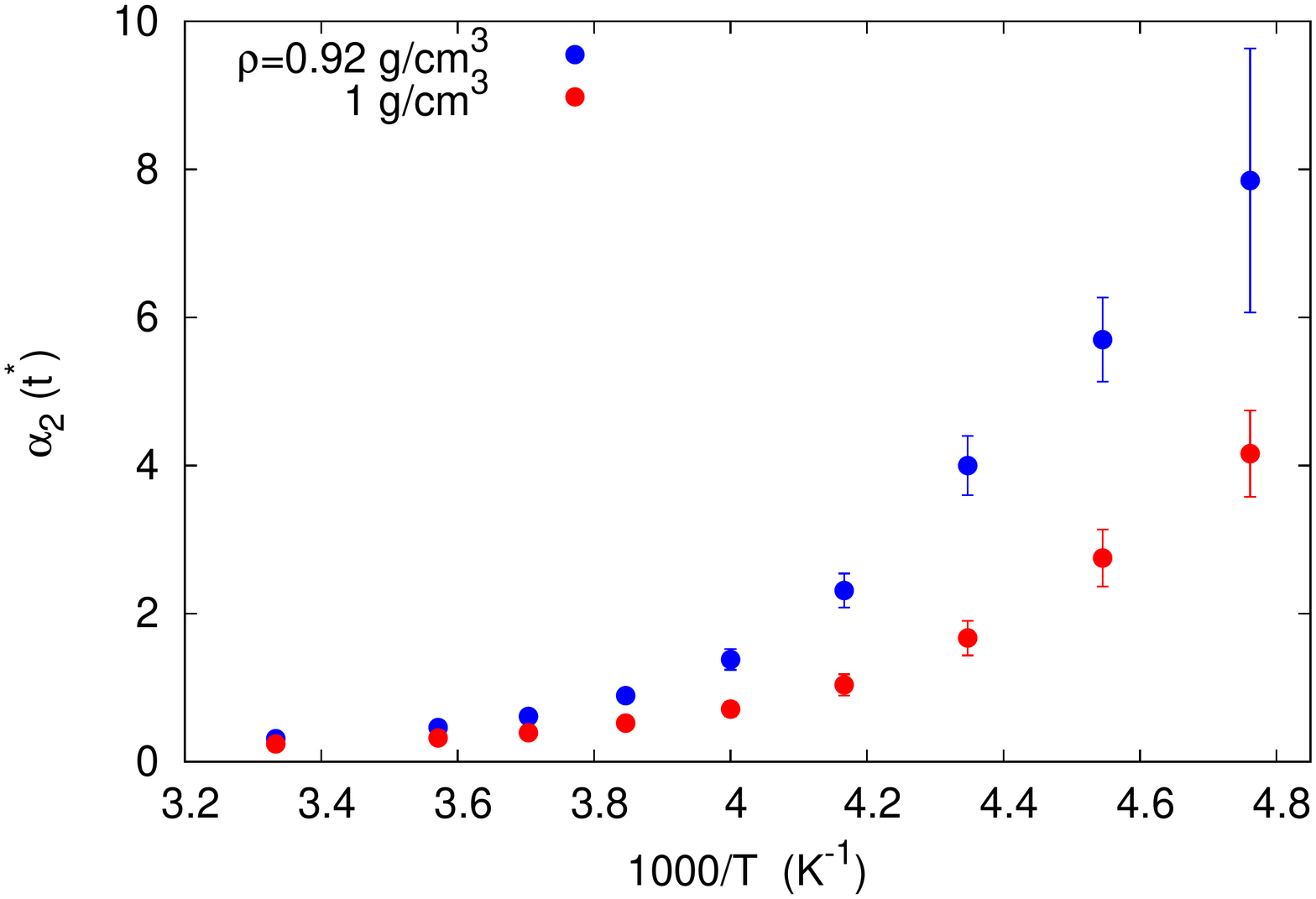}
\caption{(color online) Maximum of the Non Gaussian parameter $\alpha_{2}(t^{*})$ versus the inverse of the temperature $T$.
The curves show the increase of the cooperative motions when the temperature drops.
Note that the cooperative motions are larger for the smaller density as expected from the deviation of the diffusion coefficient to an Arrhenius law in Figure 3.
} 
\label{f6}
\end{figure}

Figure \ref{f6} shows the evolution of the maximum value of the Non-Gaussian parameter $\alpha_{2}(t^{*})$.
As discussed above, the NGP measures  the intensity of cooperative motions in the liquid.
We see on the Figure the rapid increase of the NGP (i.e. cooperative motions) when the temperature drops.
The Figure also shows that the cooperative motions are larger and increase faster at low density.

\vskip 0.5cm
To resume that first part, supercooled water \cite{bulk,confine,mixture,pressure} displays large cooperative motions in comparison to other liquids\cite{compare,silica,finite1,model}.
We observe here that behavior from the large fragility of water (Figure \ref{f3}) and from the Non-Gaussian parameters (Figure \ref{f6}).

How does these particularly large cooperative motions relate to the high temperature particularities of water (the hydrogen bonding and the possible liquid-liquid transition in the no man's land) ?
A first possible explanation is that the existence of polyamorphism favors structural fluctuations in the supercooled liquid that result in large dynamic heterogeneity.
If that picture is correct, we expect the cooperative motions to sharply increase near the liquid-liquid transition between LDL and HDL, as well as in mixtures\cite{mixture}.
Actually the cooperative motions increase in low temperature water when we decrease the density from $\rho_{0}=1g/cm^{3}$ to $\rho_{1}=0.92g/cm^{3}$ to promote the LDL phase. During that density decrease however the viscosity  increases, a particular behavior of water, and the increase of cooperative motions could be attributed to that increase of the viscosity.
A second possible origin of these large cooperative motions is that the network structure organizes the liquid increasing the probability of string like motions.
In favor of that second explanation we observe also relatively large cooperative motions in silica\cite{silica,finite1}.
Also the string like cooperative motions follow the pre-existing structure leading to curved string motions for water.
Note that the two explanations do not exclude each other. %and are probably both present in water.

\vskip 0.5cm

\subsection{Structure related dynamic heterogeneities}

Most scientists expect cooperative motions in supercooled liquids to be related to some structural defects. 
The underlying idea being that if the cage surrounding a molecule is too tight, the molecule is less likely to move.
Similarly if the cage is loose the molecule will move more likely than the average.
This leads to the hypothesis of structural fluctuations known as hard zones and soft zones that result to the least mobile and most mobile molecules aggregation.
However if there is some structure related effect indirectly observed from the propensity\cite{asaph1,asaph2}, the structural modification leading to these effects is elusive in most liquids. 

Contrary to other liquids, in supercooled water we find a clear relation between the local structure and the cooperative motions\cite{bulk}.
As discussed above, we actually find for water that the local structure around moving molecules is less organized than around non moving molecules.  {\color{black} We observe this behavior in Figure 1b of ref. \cite{bulk} from a comparison between the radial distribution functions around mobile, non-mobile, and average molecules.}
Why is that structural modification  visible in water and not in other liquids ?\\
Here again polyamorphism induced by the existence of different minima in the potential energy, 
%that is the existence of two 'stable' or 'metastable' but quite different structures 
will increase the structural fluctuations, leading to a more visible correlation between structure and dynamics. %Notice that the correlation between structure and dynamics is also seen at high temperature as the maximum of density is correlated to a maximum of the viscosity. 

%\vskip 0.5cm
%\subsection{ A network glass former}

%The network structure induced by the hydrogen bonds continues to exist at low temperature when the liquid is supercooled.
%The characteristic time for the bond reorganization however increases as the temperature drops.

\vskip 0.5cm
\subsection{ Confinement inside nanopores}

Confinement of liquid water inside nano pores permits to avoid crystallization at $T_{H}$, and thus gives experimental access to the no man's land. 
For that reason among others, the effects of confinement of supercooled water has been the subject of a large number of works\cite{conf1,conf2,conf3,conf4,conf5,conf6,conf7}. However the fact that confined supercooled water is or not  significantly different from bulk supercooled water is still a matter of debate.
In supercooled water, confinement increase or decrease the viscosity  depending on the wall hydrophilic or hydrophobic nature.
Confinement is also expected to cutoff the cooperative motions as they cannot propagate inside the wall. The diameter of the pore thus appears as a maximum cooperative length for two dimensions of the system.
Simulations however show that nature is more complex.
Actually, for an hydrophilic wall, the cooperative motions do not decrease but instead increase when supercooled water is confined\cite{confine}.

\vskip 1cm
\section{Conclusion}

 Water is a complex liquid due to its polyamorphism (i.e. the possibility to have different structures) and hydrogen bonding that leads to a network structure.
These characteristics lead to various anomalies that extend to the supercooled region. Inside that category we have the density and viscosity anomalous behaviors.
Moreover some new types of anomalies appear at low temperatures, that are mainly amplified behaviors of supercooled liquids.
In that second category we have the large cooperative motions and a visible connection between dynamic heterogeneities and the underlying structure of the liquid.

%\vskip 0.5cm
%{\color{black} \bf Open questions}:\\

\end{document}